# On separation of time scales in pharmacokinetics

Piekarski S, Rewekant M.
IPPT PAN, WUM

**Abstract**

A lot of criticism against the standard formulation of pharmacokinetics has been raised by several authors. It seems that the natural reaction for that criticism is to comment it from the point of view of the theory of conservation laws.
Simple example of balance equations for the intravenous administration of drug has been given in 2011 and the corresponding equations for extravasal administration are in the text. In principle, the equations of that kind allow one to describe in the self – consistent manner different processes of administration, distribution, metabolism and elimination of drugs. Moreover, it is possible to model different pharmacokinetic parameters of the non-compatmental pharmacokinetics and therefore to comment Rosigno's criticism. However, for practical purposes one needs approximate methods, in particular, those based on separation of the time scales. In this text, such method is described and its effectiveness is discussed. Basic equations are in the next chapter. Final remarks are at the end of the text.

# On approximate solutions in pharmacokinetics

In order to generalize system of equations describing intravenous administration [6] to the case of extravenous administration one has to add the additional term $Q(t)$ describing the rate of absorption of a drug in the plasma [5].
Moreover, in this text the processes of elimination of a drug from the central compartment shall be described by the "arbitrary" function $f[s(t)]$ instead of the linear expression $-\propto s(t)$ (this modification is not very important since most drugs are eliminated according to the „linear" formula).

Therefore, we start with the following equations:

$$\frac{\partial s(t)}{\partial t} = -k_+ e(t)s(t) + k_- c(t) - f[s(t)]$$
$$+ \frac{\beta}{V_0}[n(t) - s(t)] + Q(t)$$

(1)

$$\frac{\partial e(t)}{\partial t} = -k_+ e(t)s(t) + k_- c(t)$$

(2)

$$\frac{\partial c(t)}{\partial t} = k_+ e(t)s(t) - k_- c(t)$$

(3)

$$\frac{\partial}{\partial t} V_{eff} n(t) = -\beta[n(t) - s(t)],$$

(4)

where s(t) denotes free drug concentration in the central compartment, $s(t) \geq 0$,
$e(t), e(t) \geq 0$ denotes protein concentration,
$c(t), c(t) \geq 0$ denotes the concentration of the drug – protein complex,
$k_+$ i $k_-$ are the reaction rates („on" and „off" rates, $k_+ > 0$, $k_- > 0$).
$N(t)$ is the quantity of a drug present in the peripheral compartment at the time instant $t$,
$V_{eff}$ is the volume of the peripheral compartment, $n(t)$ denoters the concentration of a free drug in the peryphecic compartment,

$$N(t) = V_{eff} n(t)$$

(5)

$V_0$ denotes the volume of the central compartment (usually 5 or 6 litres).
The total quantity of a drug in the time instant $t$ therefore reads

$$N_{tot}(t) = V_0[c(t) + s(t)] + V_{eff} n(t).$$

(6)

After comparing (2) and (3) one can see that the sum

$$c(t) + e(t)$$

(7)

is a constant of motion which shall be denoted as $\Lambda$:

$$\Lambda = c(t) + e(t) \geq 0.$$

(8)

After insertion

$$e(t) = \Lambda - c(t) \tag{9}$$

into (1) - (4) one arrives at

$$\frac{\partial s(t)}{\partial t} = -k_+[\Lambda - c(t)]s(t) + k_-c(t) - f[s(t)] + \frac{\beta}{V_0}[n(t) - s(t)] + Q(t) \tag{10}$$

$$\frac{\partial}{\partial t}[\Lambda - c(t)] = -k_+[\Lambda - c(t)]s(t) + k_-c(t) \tag{11}$$

$$\frac{\partial c(t)}{\partial t} = k_+[\Lambda - c(t)]s(t) - k_-c(t) \tag{12}$$

$$\frac{\partial}{\partial t}V_{eff}n(t) = -\beta[n(t) - s(t)] \tag{13}$$

and

$$\frac{\partial s(t)}{\partial t} = -k_+[\Lambda - c(t)]s(t) + k_-c(t) - f[s(t)] + \frac{\beta}{V_0}[n(t) - s(t)] + Q(t) \tag{14}$$

$$-\frac{\partial}{\partial t}c(t) = -k_+[\Lambda - c(t)]s(t) + k_-c(t) \tag{15}$$

$$\frac{\partial c(t)}{\partial t} = k_+[\Lambda - c(t)]s(t) - k_-c(t) \tag{16}$$

$$\frac{\partial}{\partial t}V_{eff}n(t) = -\beta[n(t) - s(t)]. \tag{17}$$

It is worth to stress that the invariant $\Lambda$ is not a „dynamical variable" and equations (15) and (16) are dependent (they can be transformed one into another after multiplication with -1) and therefore we are left with the following system of three equations:

$$\frac{\partial s(t)}{\partial t} = -k_+[\Lambda - c(t)]s(t) + k_-c(t) - f[s(t)] + \frac{\beta}{V_0}[n(t) - s(t)] + Q(t) \tag{18}$$

$$\frac{\partial c(t)}{\partial t} = k_+[\Lambda - c(t)]s(t) - k_-c(t) \tag{19}$$

$$\frac{\partial}{\partial t} V_{eff} n(t) = -\beta[n(t) - s(t)].$$

(20)

In order to discuss the problem of time scales for the system (18) – (20) one should take into account that – according to the pharmacokinetic literature – the processes of drug – protein binding occur in the time – scales of miliseconds. Let us start our discusion with writing equations (1) – (4) in the following equivalent form:

$$\frac{\partial s(t)}{\partial t} = -k_- \left[\frac{k_+}{k_-} e(t)s(t) - c(t)\right] - f[s(t)] + \frac{\beta}{V_0}[n(t) - s(t)] + Q(t)$$

(21)

$$\frac{\partial e(t)}{\partial t} = -k_- \left[\frac{k_+}{k_-} e(t)s(t) - c(t)\right]$$

(22)

$$\frac{\partial c(t)}{\partial t} = k_- \left[\frac{k_+}{k_-} e(t)s(t) - c(t)\right]$$

(23)

$$\frac{\partial}{\partial t} V_{eff} n(t) = -\beta[n(t) - s(t)].$$

(24)

The Guldberg - Waage law states that the condition of temporary chemical equilibrium between the concentrations of the complex $c(t)$ and the concentrations of substrates $e(t)$ and $s(t)$ is

$$\frac{k_+}{k_-} e(t)s(t) = c(t).$$

(25)

Therefore, the state of chemical equilibrium is characterized by

$$\frac{k_+}{k_-} = K_a$$

(26)

while the relaxation of the system towards chemical equilibrium is characterized by value of the parameter $k_-$.
Now we want to determine the form of our equations in the limit

$$\frac{k_+}{k_-} = const, \quad k_- \to \infty$$

(27)

without entering in rigorous derivations (some control on the validity of this approximation can be obtained by numerical experiments): in the limit (27) one

can assume that all quantities in cemtral compartment are in a temporary chemical equilibrium

$$\frac{k_+}{k_-}e(t)s(t) - c(t) = 0.$$

(28)

However, such an approximation has to take into account the exact identity (8) and therefore it is better to start with writing equations (18) – (20)

$$\frac{\partial s(t)}{\partial t} = -k_-[\Lambda - c(t)]s(t) + k_-c(t) - f[s(t)] + \frac{\beta}{V_0}[n(t) - s(t)] + Q(t)$$

(29)

$$\frac{\partial c(t)}{\partial t} = k_-[\Lambda - c(t)]s(t) - k_+c(t)$$

(30)

$$\frac{\partial}{\partial t}V_{eff}n(t) = -\beta[n(t) - s(t)]$$

(31)

in the form

$$\frac{\partial s(t)}{\partial t} = -k_-\left\{\frac{k_+}{k_-}[\Lambda - c(t)]s(t) - c(t)\right\} - f[s(t)] + \frac{\beta}{V_0}[n(t) - s(t)] + Q(t)$$

(32)

$$\frac{\partial c(t)}{\partial t} = k_-\left\{\frac{k_+}{k_-}[\Lambda - c(t)]s(t) - c(t)\right\}$$

(33)

$$\frac{\partial}{\partial t}V_{eff}n(t) = -\beta[n(t) - s(t)].$$

(34)

In order to get more insight in the nature of equations (32) – (34) in the limit (27) let us write

$$k_- = \frac{1}{\varepsilon}$$

(35)

and then

$$\frac{\partial s(t)}{\partial t} = -\frac{1}{\varepsilon}\left\{\frac{k_+}{k_-}[\Lambda - c(t)]s(t) - c(t)\right\} - f[s(t)] + \frac{\beta}{V_0}[n(t) - s(t)] + Q(t)$$

(36)

$$\frac{\partial c(t)}{\partial t} = \frac{1}{\varepsilon}\left\{\frac{k_+}{k_-}[\Lambda - c(t)]s(t) - c(t)\right\}$$

(37)

$$\frac{\partial}{\partial t} V_{eff} n(t) = -\beta[n(t) - s(t)].$$

(38)

One can see that the perturbation introduces by small $\varepsilon$ is singular and here we just quess the final form of the limit equations taking into account the constraint imposed by fixed value of the invariant $\Lambda$ for the condition of the temporal chemical equilibrium

$$\frac{k_+}{k_-}[\Lambda - c(t)]s(t) - c(t) = 0.$$

(39)

Those equations are

$$\frac{\partial s(t)}{\partial t} = -f[s(t)] + \frac{\beta}{V_0}[n(t) - s(t)] + \varepsilon_s\left[s(t), \Lambda, \frac{k_+}{k_-}\right] Q(t)$$

(40)

$$\frac{\partial}{\partial t} V_{eff} n(t) = -\beta[n(t) - s(t)]$$

(41)

and can bo solved for $s(t)$ i $n(t)$ while $c(t)$ can by determined from the condition of the temporal chemical equilibrium.

Additionally, the initial condition has to be correspondingly changed and this change shall be discussed later by means of a simple example.

The symbol $\varepsilon_s\left[s(t), \Lambda, \frac{k_+}{k_-}\right]$ in (40) denotes the fraction of the free form of a drug in central comparetment (which is represented here as a function of quantities $s(t), \Lambda, \frac{k_+}{k_-}$).

In order to determine the explicit expression for $\varepsilon_s\left[s(t), \Lambda, \frac{k_+}{k_-}\right]$ let us consider first two arbitrary non – negative quantities s(t) i c(t) and let

$$\varepsilon_s(t) = \frac{s(t)}{s(t) + c(t)}$$

(42)

and

$$\varepsilon_c(t) = \frac{c(t)}{s(t) + c(t)}.$$

(43)

After dividing simultaneously the nominator and denominator in (42) by s(t) one arives at

$$\varepsilon_s(t) = \frac{1}{1 + \frac{c(t)}{s(t)}}.$$

In order to arrive at the explicit formula for the quotient $\frac{c(t)}{s(t)}$ for Guldberg – Waage formula for temporary chemical equilibrium one can apply the identities

$$\frac{k_+}{k_-} = K_a, \quad K_a e(t) s(t) = c(t). \tag{44}$$

The identity (8) can be transformed to the form

$$e(t) = \Lambda - c(t) \tag{45}$$

and (46) can be inserted into (45) what results in

$$K_a [\Lambda - c(t)] s(t) = c(t) \tag{46}$$

and later

$$K_a \Lambda s(t) - K_a c(t) s(t) = c(t). \tag{47}$$

Therefore

$$K_a \Lambda s(t) = c(t) + K_a c(t) s(t), \tag{48}$$

$$K_a \Lambda s(t) = c(t)[1 + K_a s(t)] \tag{49}$$

later

$$c(t) = \frac{K_a \Lambda s(t)}{[1 + K_a s(t)]} \tag{50}$$

and finally

$$\frac{c(t)}{s(t)} = \frac{K_a \Lambda}{[1 + K_a s(t)]}. \tag{51}$$

After inserting (52) into (44) one obtains

$$\varepsilon_s(t) = \frac{1}{1 + \frac{c(t)}{s(t)}} =$$

$$\cfrac{1}{1+\cfrac{K_a\Lambda}{[1+K_as(t)]}} = \cfrac{1}{\cfrac{1+K_as(t)}{1+K_as(t)} + \cfrac{K_a\Lambda}{[1+K_as(t)]}} =$$

$$\cfrac{1}{\cfrac{1+K_as(t)+K_a\Lambda}{1+K_as(t)}} = \cfrac{1+K_as(t)}{1+K_as(t)+K_a\Lambda}$$

$$\cfrac{1}{\cfrac{1+K_as(t)+K_a\Lambda}{1+K_as(t)}} = \cfrac{1+K_as(t)}{1+K_as(t)+K_a\Lambda}.$$

(53)

Now all parts of the system (40) – (41) are explicitly known and this system can be solved for $s(t)$ and $n(t)$. Later $c(t)$ can be determined from the identity (51). Let us illustrate this scheme on a simple example.

The system (40),(41) describes extravenous administration of a drug (derscribed by the term $Q(t)$) and allows the "nonlinear" elimination elimination of a drug. It is well – known that most drugs are eliminated according to the „linear"expression [1]:

$$f[s(t)] = \alpha s(t).$$

(54)

For the case of intravenous administration and fot the „linear"elimination our system reduces to

$$\frac{\partial s(t)}{\partial t} = -\alpha s(t) + \frac{\beta}{V_0}[n(t) - s(t)]$$

(55)

$$\frac{\partial}{\partial t}V_{eff}n(t) = -\beta[n(t) - s(t)].$$

(56)

The initial conditions are of the form

$$s(t = 0) > 0$$

(57)

$$n(t = 0) = 0$$

(58)

Those equations can be easily solved but we assume further simplifications

$$V_{eff} = V_0 = 1$$

(59)

leading to

$$\frac{\partial s(t)}{\partial t} = -\alpha s(t) + \beta [n(t) - s(t)] \tag{60}$$

$$\frac{\partial}{\partial t} n(t) = -\beta [n(t) - s(t)]. \tag{61}$$

The solution of the system (60),(61) is

$$s(t) = \frac{1}{2} x_0 e^{\left[-\frac{\alpha}{2} - \beta + \frac{\sqrt{\Delta}}{2}\right]t} + \frac{1}{2} x_0 e^{\left[-\frac{\alpha}{2} - \beta - \frac{\sqrt{\Delta}}{2}\right]t} \tag{62}$$

$$n(t) = \frac{1}{2} x_0 \frac{\sqrt{\Delta} + \alpha}{2\beta} \left\{ e^{\left[-\frac{\alpha}{2} - \beta + \frac{\sqrt{\Delta}}{2}\right]t} - e^{\left[-\frac{\alpha}{2} - \beta - \frac{\sqrt{\Delta}}{2}\right]t} \right\} \tag{63}$$

$$\Delta = 4\beta^2 + \alpha^2 \tag{63}$$

One can see that (62),(63) for t=0 give

$$s(t = 0) = \frac{1}{2} x_0 + \frac{1}{2} x_0 = x_0 \tag{64}$$

$$n(t = 0) = 0 \tag{65}$$

After inserting (62) into (51) one can determine the corresponding concentration of the complex

$$c(t) = \frac{K_a \Lambda \left\{ \frac{1}{2} x_0 e^{\left[-\frac{\alpha}{2} - \beta + \frac{\sqrt{\Delta}}{2}\right]t} + \frac{1}{2} x_0 e^{\left[-\frac{\alpha}{2} - \beta - \frac{\sqrt{\Delta}}{2}\right]t} \right\}}{\left[1 + K_a \left\{ \frac{1}{2} x_0 e^{\left[-\frac{\alpha}{2} - \beta + \frac{\sqrt{\Delta}}{2}\right]t} + \frac{1}{2} x_0 e^{\left[-\frac{\alpha}{2} - \beta - \frac{\sqrt{\Delta}}{2}\right]t} \right\}\right]} \tag{66}$$

($\Delta$ is defined by (63)).
For t = 0, the concentration of the complex reads

$$c(t = 0) = \frac{K_a \Lambda \left\{ \frac{1}{2} x_0 + \frac{1}{2} x_0 \right\}}{\left[1 + K_a \left\{ \frac{1}{2} x_0 + \frac{1}{2} x_0 \right\}\right]} = \frac{K_a \Lambda x_0}{[1 + K_a x_0]}$$

Therefore, in our model the total concentration of a drug for t=0 is (67)

$$s(t=0) + c(t=0) = x_0 + \frac{K_a \Lambda x_0}{[1+K_a x_0]}$$

(68)

and the dose of a drug absorbed in a single intravascular administration for t=0 is

$$V_0 \left[ x_0 + \frac{K_a \Lambda x_0}{[1+K_a x_0]} \right].$$

(69)

It should be stressed that in rigorous model (describing the states outside the chemical equilibrium) the dose absorbed during a single intravenous administration is given only in terms of the free form of a drug.
The form of the expression (68) is a result of our approximate scheme of solution.
One can see that (64) becomes the sum of exponential functions only in the limit

$$K_a s(t) \ll 1$$

(70)

that is, for

$$s(t) \ll \frac{1}{K_a}$$

(71)

The formula (59) can be equivalently written also in other forms

$$c(t) = \frac{\Lambda s(t)}{\left[\frac{1}{K_a} + s(t)\right]}$$

(72)

and

$$c(t) = \frac{s(t)}{\left[\frac{1}{\Lambda K_a} + \Lambda s(t)\right]}$$

(73)

which can be convenient for a more detailed discussion.

## Final remarks

In standard pharmacokinetics one usually assumes that the solutions for the drug concentrations are the sums of the exponential functions [1,8]. However, it is not so in our model and the consistency with literature takes place in the "linear regime"only.

In seems important to make numerical simulations of our approximation scheme. We hope that this scheme requires further discussion and that it can be useful for finding approximate solutions of pharmacokinetic problems.